\documentclass[aps,prl,preprint]{revtex4}

\usepackage{epsfig}
\usepackage{amssymb}
\usepackage{times}

\begin{document}

{\Large  DESY 22-180}

{\Large  November 2022}

\bigskip
\bigskip
\bigskip
\bigskip


\title{Two-color operation of a soft x-ray FEL with alternation of undulator tunes}



\author{E.A. Schneidmiller}
\email[]{evgeny.schneidmiller@desy.de}
\affiliation{Deutsches Elektronen-Synchrotron DESY, Notkestr. 85, 22607 Hamburg, Germany}
\author{I.J. Bermudez Macias} 
\affiliation{Deutsches Elektronen-Synchrotron DESY, Notkestr. 85, 22607 Hamburg, Germany}
\author{M. Beye}
\affiliation{Deutsches Elektronen-Synchrotron DESY, Notkestr. 85, 22607 Hamburg, Germany} 
\author{M. Braune} 
\affiliation{Deutsches Elektronen-Synchrotron DESY, Notkestr. 85, 22607 Hamburg, Germany}
\author{M.-K. Czwalinna}
\affiliation{Deutsches Elektronen-Synchrotron DESY, Notkestr. 85, 22607 Hamburg, Germany}
\author{S. D\"usterer}
\affiliation{Deutsches Elektronen-Synchrotron DESY, Notkestr. 85, 22607 Hamburg, Germany}
\author{B. Faatz} 
\affiliation{Shanghai Advanced Research Institute, Chinese Academy of Sciences, Haike Road 99, Shanghai 201210, China}
\author{R. Ivanov}
\affiliation{Deutsches Elektronen-Synchrotron DESY, Notkestr. 85, 22607 Hamburg, Germany} 
\author{F. Jastrow}
\affiliation{Deutsches Elektronen-Synchrotron DESY, Notkestr. 85, 22607 Hamburg, Germany}
\author{M. Kuhlmann}
\affiliation{Deutsches Elektronen-Synchrotron DESY, Notkestr. 85, 22607 Hamburg, Germany}
\author{J. Roensch-Schulenburg} 
\affiliation{Deutsches Elektronen-Synchrotron DESY, Notkestr. 85, 22607 Hamburg, Germany}
\author{S. Schreiber}
\affiliation{Deutsches Elektronen-Synchrotron DESY, Notkestr. 85, 22607 Hamburg, Germany} 
\author{A. Sorokin}
\thanks{This paper is dedicated to the memory of Andrey Sorokin.}
\affiliation{Deutsches Elektronen-Synchrotron DESY, Notkestr. 85, 22607 Hamburg, Germany} 
\author{K. Tiedtke} 
\affiliation{Deutsches Elektronen-Synchrotron DESY, Notkestr. 85, 22607 Hamburg, Germany}
\author{M.V. Yurkov}
\affiliation{Deutsches Elektronen-Synchrotron DESY, Notkestr. 85, 22607 Hamburg, Germany}
\author{J. Zemella}
\affiliation{Deutsches Elektronen-Synchrotron DESY, Notkestr. 85, 22607 Hamburg, Germany}

\date{\today}

\begin{abstract}
FLASH is the first soft X-ray FEL user facility, routinely providing brilliant photon beams for users since 2005.
The second undulator branch of this facility, FLASH2, is gap-tunable which allows to test and use advanced
lasing concepts. In particular, we developed a two-color operation mode based on the alternatingly tuned
undulator segments (every other segment is tuned to the second wavelength). This scheme
is advantageous in comparison with a subsequent generation of two colors in two consecutive sections of the
undulator line. First, source positions of the two FEL beams are close to each other which makes it easier to
focus them on a sample. Second, the amplification is more efficient in this configuration since the segments with
respectively "wrong" wavelength still act as bunchers. We studied operation of this scheme in the regime of small and large separation of tunes (up to a factor of two).
We developed new methods for online intensity measurements
of the two colors simultaneously that require a combination of two detectors. We also demonstrated our capabilities to measure 
spectral and temporal properties of two pulses with different wavelengths. 
\end{abstract}

\maketitle

\section{Introduction}

Two-color lasing is a popular operation mode of X-ray FEL user facilities. A possible way to generate two
colors in a gap-tunable undulator, operating in SASE regime, is to split the undulator into two sections, and
set different K-values in
each of these sections. An additional useful element for X-ray pump, X-ray probe experiments could be a chicane between
the two sections to control a time delay between the pulses of two different colors. One can use the same electron bunch
(parts of the same bunch) for lasing in the two sections of the undulator \cite{hara}. In this case the beam quality
is partially spoiled in the first section (that typically operates at the onset of saturation), and one has to
compromise intensities of two colors. As an alternative, one can use fresh-bunch technique based on the
principle of
the betatron switcher \cite{brinkmann}. A practical realization of this principle was done in a way
that the head and the tail of the bunch lase in two different sections of the undulator \cite{lutman,guetg}.

One of the issues with these sectioned-undulators schemes (or, split undulators) is a significant
spatial separation of effective source positions
of the two X-ray beams. This separation can make it difficult to efficiently focus both beams on a
sample. As an alternative, one can consider two-color lasing in the undulator with the alternation of
tunes of single undulator segments enabling near positions of source points. This scheme was realized at LCLS for small separation of tunes (about $1\%$) and
was described in \cite{gain-modul} as the gain-modulated FEL.

We studied extensively the two-color mode with the alternation of undulator tunes at FLASH \cite{flash,fl2-njp} in 2018-2021 in the
regime of a large separation of tunes spanning up to a factor of two. We developed new methods of photon diagnostics for this operation mode, including simultaneous measurements of intensity, spectral and temporal properties. The main results are presented below in this paper.

\begin{figure}[b]

\includegraphics[width=.8\textwidth]{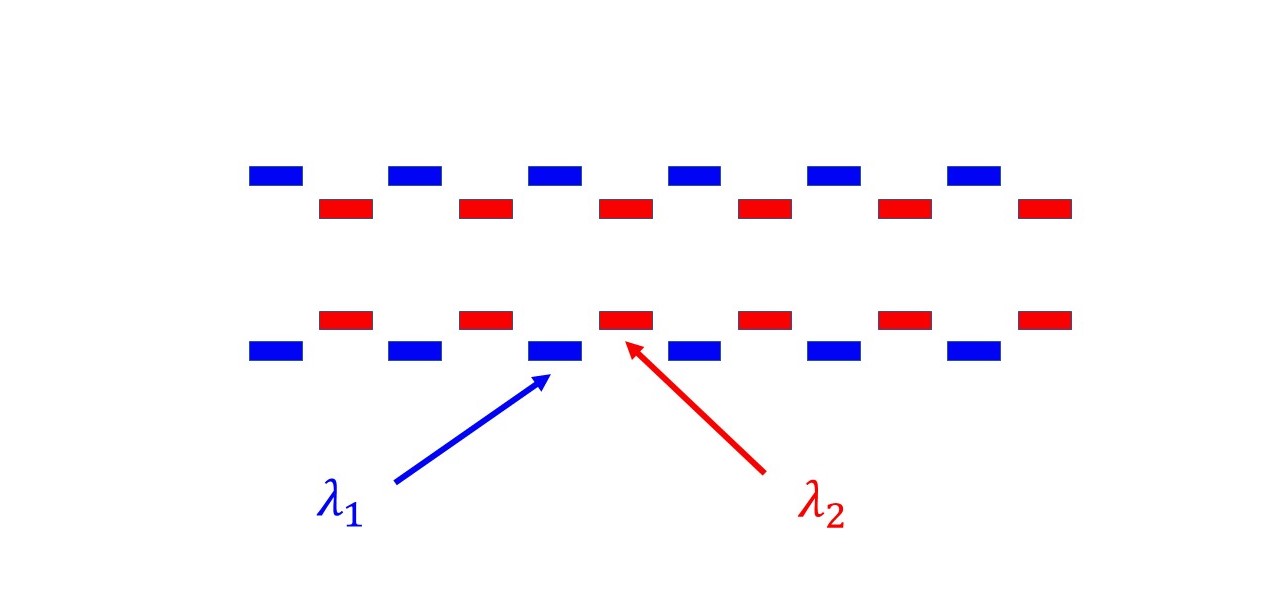}

\caption{Conceptual scheme of two-color lasing with the alternation of undulator tunes.}

\label{scheme-2-colors}
\end{figure}

\section{Two-color lasing with the alternation of undulator tunes}

Let us consider gap-tunable segmented undulators that are normally used at X-ray FEL facilities.
For generation of two colors, all odd segments of an undulator are tuned to
wavelength $\lambda_1$, and even segments to $\lambda_2$ (see Fig.~\ref{scheme-2-colors}).
With respect to the amplification of the electromagnetic wave with
the wavelength $\lambda_1$, the FEL process is disrupted as soon as the electron beam leaves a $\lambda_1$
segment and enters a "wrong" segment tuned to $\lambda_2$. However, energy modulations in the electron bunch,
accumulated due to the
interaction with the electromagnetic field in the $\lambda_1$ segment, continue to get converted into density
modulations (bunching) in the $\lambda_2$ segment due to its longitudinal dispersion.
Thus, in the next $\lambda_1$ segment the beam with an enhanced bunching
quickly radiates a stronger field than the one coming from the previous $\lambda_1$ segment (which in
addition is diffracted), and the FEL process
continues with higher amplitudes. In some sense the mechanism is similar to that of the multi-stage (or
distributed) optical klystron \cite{vinokurov,litvinenko,ssy-ok,huang-ok}.
In this qualitative description we do not
consider effects of longitudinal velocity spread due to the energy spread and emittance of the electron beam.
According to our estimates, these effects were practically negligible in our experiments.

The longitudinal dispersion of an undulator segment is characterized by a transfer matrix element
$R_{56} = 2 N_w \lambda$, where $N_w$ is the number of undulator periods per segment and $\lambda$ is
the resonance wavelength. If $\lambda_1 < \lambda_2$, the FEL gain is, obviously, weaker for $\lambda_1$ in the
corresponding $\lambda_1$ segments. However, the $R_{56}$ in $\lambda_2$ segments is stronger and gives a larger addition to
the final gain at $\lambda_1$ wavelength. Thus, the total gain in the linear regime
can be comparable even if $\lambda_1$ is significantly shorter than $\lambda_2$. In principle, the number of
segments does not have to be the same for $\lambda_1$ and $\lambda_2$, this depends on desired intensity ratio. 
If the undulator line is sufficiently long, nonlinear effects start to play a role in the last segments. In this
case the amplification processes at both wavelengths are not independent anymore, and they start to compete in
terms of modification of the longitudinal phase space of the electron beam. As a result, the radiation power is
somewhat smaller compared to standard single-color lasing in saturation regime. However, it can still be
sufficient for many experiments.

Finally, we should note that the considered scheme was the only possibility to establish two-color operation of FLASH because the undulator is too short to have a sufficient gain in the split-undulator configuration. 

\section{Experimental studies at FLASH}

\begin{figure}[b]

\includegraphics[width=1.0\textwidth]{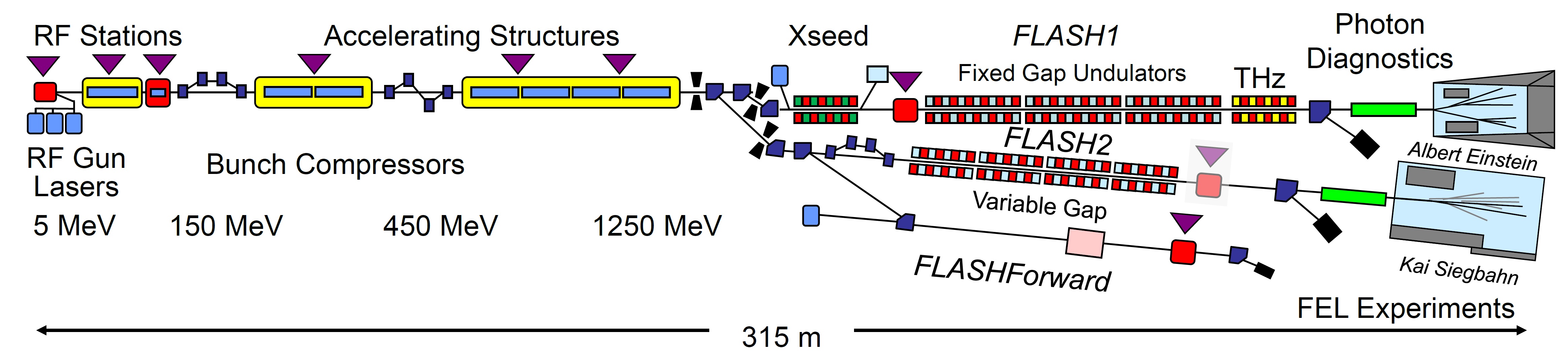}

\caption{Layout of FLASH facility}

\label{flash-layout}
\end{figure}

FLASH (Free-electron LASer in Hamburg) \cite{flash} started user operation in summer 2005 as the first free-electron laser for XUV and soft X-ray radiation. It is operated in the "self-amplified spontaneous emission" (SASE) mode \cite{kond-sal} and currently covers a wavelength range from 4 nm to about 90 nm in the first harmonic with GW peak power and pulse durations between a few fs and 200 fs. 
The electron bunches with maximum energy of 1.25 GeV are distributed between the two branches, FLASH1 and FLASH2 \cite{fl2-njp}, see Fig.~\ref{flash-layout}. The facility operates with long pulse trains (several hundred pulses) with 10 Hz repetition rate.
Presently, the facility is being upgraded towards high repetition rate seeding in FLASH1 branch \cite{fl2020plus}.

The segmented variable-gap undulators of the FLASH2 branch have a period of 3.14 cm and can provide a maximum undulator parameter K about 2.7. FLASH2 hosts twelve 2.5 m long segments with quadrupoles in the intersections.
After successful experiments with two-color lasing, the control system of the undulator was upgraded to make the operation in this regime easier. In particular, it allows user control of two-color operation. Also, special photon diagnostics methods were developed as described below, and there is online intensity diagnostics suitable for user operation.  

Below in this paper we present exemplary results obtained during two-color operation of FLASH2 undulator in 2018 - 2021. Electron energy in these experiments was between 430 MeV and 1240 MeV, bunch charge was  typically in the range 70 - 300 pC. We generated FEL wavelengths between 6 nm and 50 nm, pulse energies in both colors were up to several tens of microjoules. The spectral separation between two colors can be set from $\sim$ 1\% to $\sim$ 100\%.   
It is possible to do rough changes of the pulse energy ratio by changing, for example, between 6 + 6 configuration of the undulator segments (as shown in Fig.~\ref{scheme-2-colors}) and 5 + 7 configuration, as well as fine tuning by slightly changing the K values of a few segments.

\section{Simultaneous measurements of two colors}

\subsection{Intensity measurement}

As it was mentioned, the two colors are not generated independently, so that one can not measure their intensities
by turning off one color and measuring the other one (and vice versa). Moreover, in case of user experiments one
should have online nondisruptive measurements. Thus, one of the goals of our studies was to develop methods
for such measurements. It is
obvious that we can determine pulse energies of each of the X-ray beams using two
linear detectors having a different response for the different wavelengths. In this case we have a system of two linear equations with two unknowns and can easily retrieve pulse energies at $\lambda_1$ and $\lambda_2$.

We have four of such detectors available in the FLASH2 tunnel and experimental hall: two gas monitor detectors
(GMDs) for measurements of absolute pulse energy \cite{xgm-1}, the online photoionization spectrometer (OPIS)
for non-invasive wavelength measurements \cite{opis}, and a micro-channel-plate (MCP) detector for pulse energy
measurements with a large dynamic range \cite{mcp}.
Note that in the following we refer to ensemble-averaged pulse energies throughout the text.

As an example, let us consider the measurement we did
with the tunnel GMD and OPIS. Note that both devices are placed in the tunnel next to each other, and there are
no propagation effects that can be different for the two X-ray beams. In the time-of-flight spectra of the OPIS,
relative intensities of the two X-ray colors can be determined since the signals of the respective
photoelectrons are separated in arrival time due to different kinetic energies. From these signal
intensities a ratio of number of photons of the two wavelengths can be evaluated.
Thus, we have the first linear equation:

\begin{equation}
N_2 = p N_1 \ ,
\label{opis}
\end{equation}

\noindent where $N_1$ and $N_2$ are unknown average photon numbers,
and $p$ is the measured coefficient. If the tunnel GMD
is set to the measurements of pulse energy at $\lambda_1$, it will show spurious pulse energy
$\tilde{\cal{E}}_{gmd}$
when the second color is present:

\begin{equation}
\tilde{\cal{E}}_{gmd} =  \hbar \omega_1 \left( N_1 + N_2 \frac{\sigma_2 \gamma_2}{\sigma_1 \gamma_1} \right) \ .
\label{gmd}
\end{equation}

\noindent Here $\sigma_{1,2}$ are the photoionization cross sections and
$\gamma_{1,2}$ are the mean charges for a given gas and the two photon energies (see, e.g. \cite{tiedtke}),
$\omega_1 = 2\pi c / \lambda_1$.
Solving Eqs.~(\ref{opis}) and (\ref{gmd}) for $N_1$ and $N_2$, we get the final result for the actual
pulse energies ${\cal{E}}_1 = \hbar \omega_1 N_1$ and ${\cal{E}}_2 = \hbar \omega_2 N_2$:

\begin{displaymath}
{\cal{E}}_1 = \frac{\tilde{\cal{E}}_{gmd}} {1 + p \frac{\sigma_2 \gamma_2}{\sigma_1 \gamma_1}} \ \ \ \ , \ \ \ \ \ \
{\cal{E}}_2 = p {\cal{E}}_1 \frac{\omega_2}{\omega_1} \ .
\end{displaymath}

\begin{figure}[tb]

\includegraphics[width=.5\textwidth]{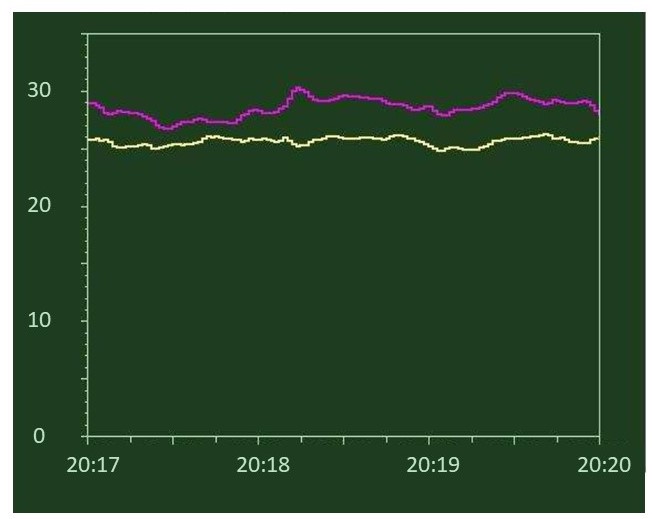}

\caption{Online measurement of FEL intensities in two-color mode with two gas monitor detectors.
Pulse energies in microjoules are shown for 7 nm (yellow) and for 10 nm (pink) versus real time.}

\label{two-gmds}
\end{figure}

In a similar way one can get simple expressions for the combination of the MCP-based detector and the tunnel GMD.
The MCP detector can be cross-calibrated with the GMD at a certain wavelength, say $\lambda_1$. When two
X-ray beams are present, it shows a linear combination of two actual pulse energies:
$\tilde{\cal{E}}_{mcp} = {\cal{E}}_1 + s {\cal{E}}_2$. It is interesting to note that in the wavelength range of our
experiments we had $s \simeq 1$. Combining this equation with Eq.~(\ref{gmd}), we obtain the actual pulse energies.

We also did the measurements with two GMDs, one in the tunnel and one in the experimental hall. They have to
be filled with different gases, then Eq.~(\ref{gmd}) can be also used for the second GMD but with the different
constants $\sigma$ and $\gamma$. Moreover, we should correct the photon numbers $N_1$ and $N_2$ in that
equation for the beamline transmission that has to be measured once for each wavelength individually. Then we
have again the system of two equations from which we obtain actual pulse energies. The two-color measurement
with two GMDs has been integrated into the GMD control software and can be used online for tuning and monitoring of this
operation mode at FLASH2 (see Fig.~\ref{two-gmds}).

\begin{figure}[tb]

\includegraphics[width=.7\textwidth]{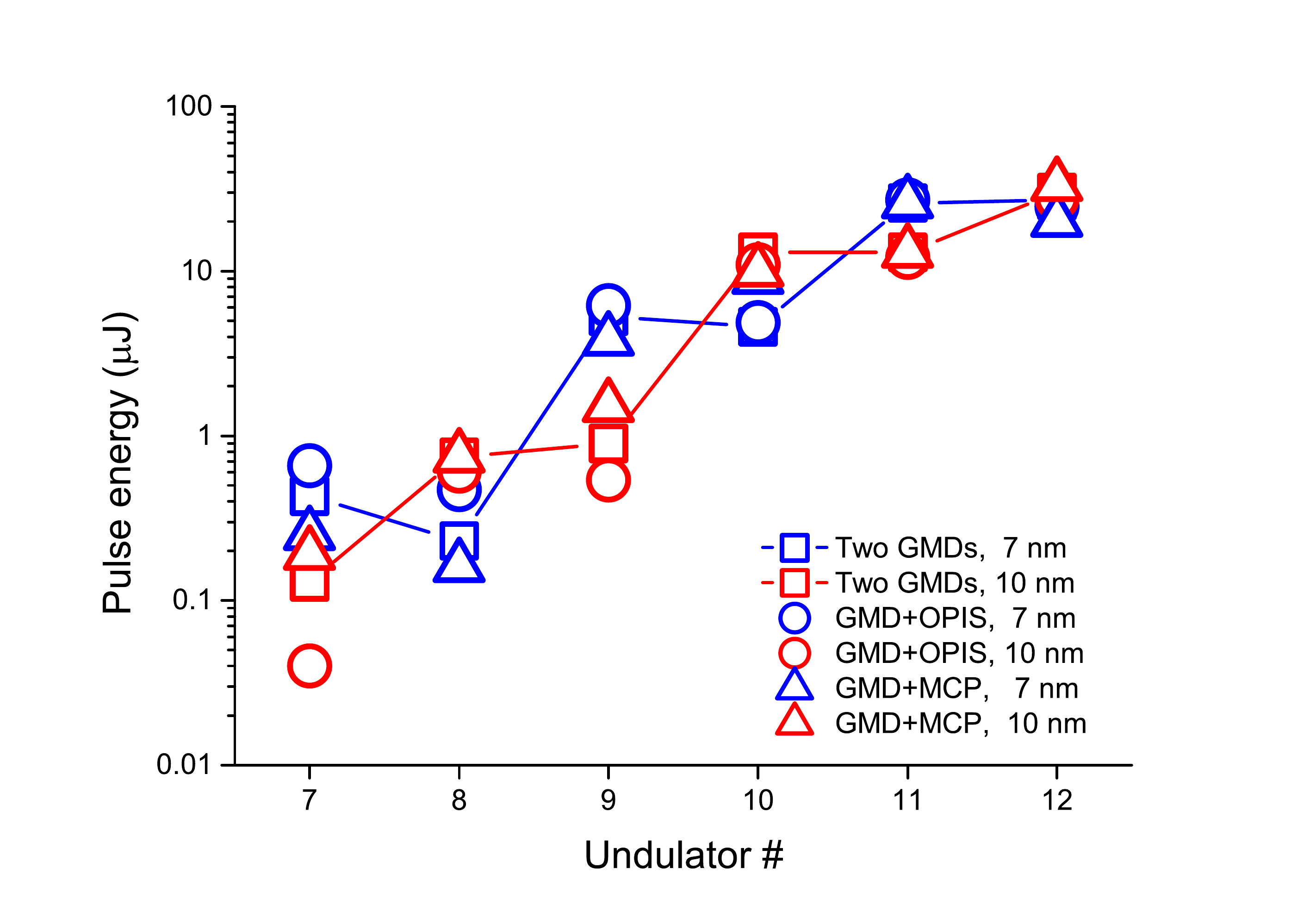}

\caption{Gain curve of two-color lasing. Pulse energies at 7 nm (blue) and at 10 nm (red) versus the undulator
segment number. Measurements were done with three methods: two GMDs (squares), GMD+OPIS (circles), and
GMD+MCP (triangles).}

\label{gain-curve-7and10}
\end{figure}

For simultaneous lasing at 7 nm and 10 nm we could measure a part of the FEL gain curve
(see Fig.~\ref{gain-curve-7and10}) using the methods described above. In general,
the three methods show a reasonable agreement at high intensity level, when the FEL operated in
a nonlinear regime and the intensities were relatively stable.

\begin{figure}[tb]

\includegraphics[width=.7\textwidth]{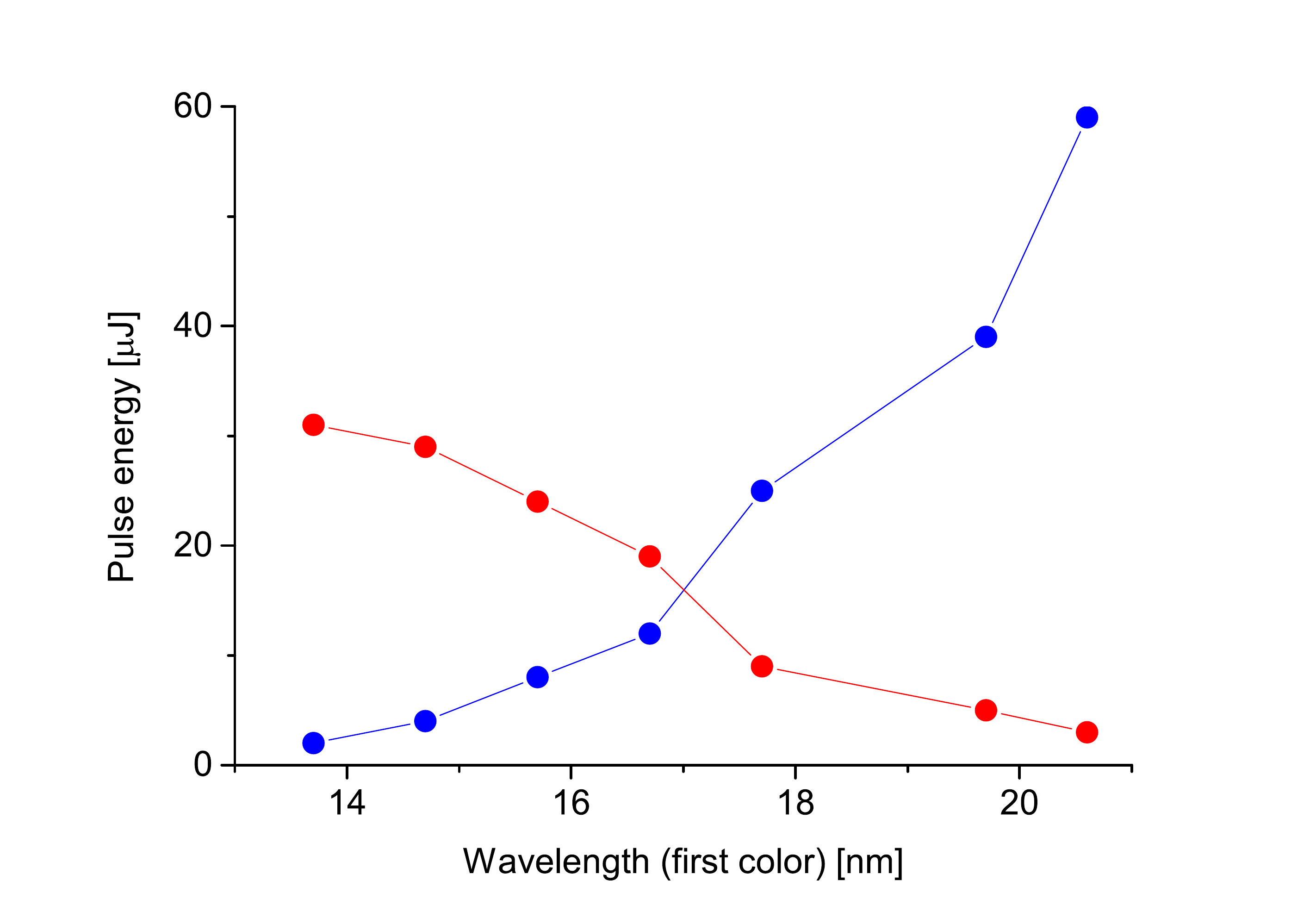}

\caption{Pulse energies of two XUV beams measured with the GMD+OPIS method. The first color (odd undulator segments)
is scanned between 13.7 nm and 20.6 nm (shown in blue). The second color (even segments) stays at 22.6 nm
(shown in red).}

\label{wl-scan}
\end{figure}

Another interesting measurement is a scan of one wavelength while
the other
stays constant. Pulse energies were measured with the help of OPIS and the tunnel GMD
(see Fig.~\ref{wl-scan}). The undulator was in
$6+6$ configuration: the even segments were kept tuned to $\lambda_2=22.6$ nm, while the
odd segments were scanned in the range  of $\lambda_1$ between 13.7 nm and 20.6 nm. One can see from
Fig.~\ref{wl-scan} that such a wavelength scan is possible in a wide range but the two colors are not generated
independently. When we increase $\lambda_1$, the pulse energy at this wavelength increases significantly at the
expense of the pulse energy reduction of $\lambda_2$ due to nonlinear effects.

\subsection{Spectral measurements}

\begin{figure}[tb]
\includegraphics[width=.5\textwidth]{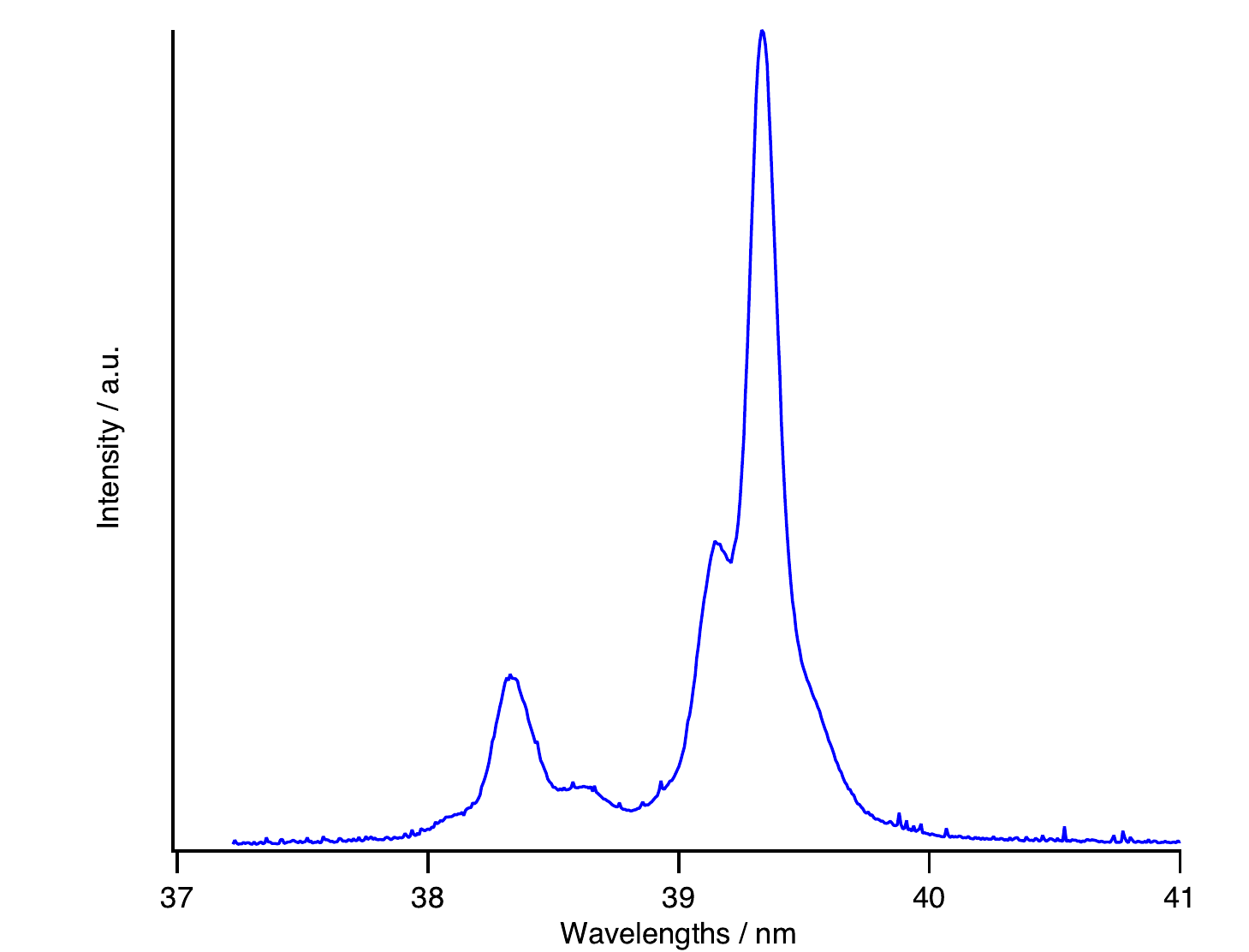}
\caption{ Sum of spectra with 3rd grating order of color 2 (12.78 nm/ 97 eV) and 4th grating order of color 1 (9.83 nm/ 126.13 eV). The photon beam intensity relation was due to experimental needs during the two-color test experiment. }
\label{spectra_sum}
\end{figure}

FLASH2 provides an invasive grating spectrometer located at the beam line FL22 providing a wavelengths range of 10 nm to 62 nm. The spectrometer is equipped with an intensified camera (IDDC) which can select one bunch of a train \cite{spectr}. Usually,  only one wavelength is monitored by the spectrometer.  Simultaneous observation of two colors may be possible with the help of higher grating orders as shown in Fig. \ref{spectra_sum}. Here, the 4th grating order of the first color at 39 nm and in the 3rd order of the second color at 38 nm are used.  Higher grating orders are preferred due to better resolution but suffer in signal to noise ratio. The first colo and the second color are recorded with 9.83 nm / 126.13 eV and 12.78 nm / 97 eV, with a bandwidth of 0.7$\%$ and 0.4$\%$.  
Since the spectrometer records the vertical beam profile the measurement can monitor the relative pointing of the two colors.  

\begin{figure}[tb]
\includegraphics[width=.5\textwidth]{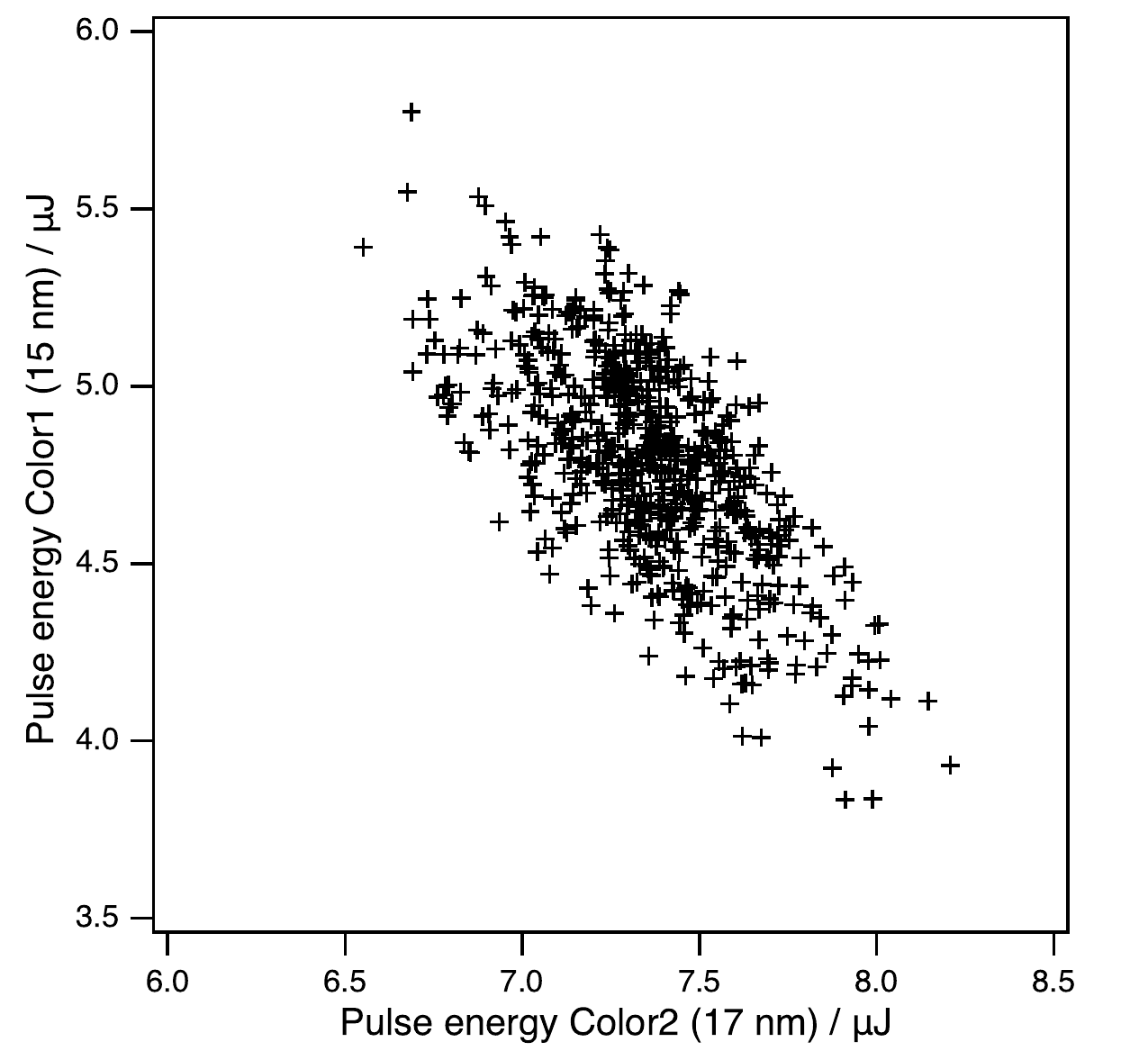}
\caption{Correlation plot of pulse energies at 15 nm and 17 nm measured with the grating spectrometer}
\label{spectr-corr}
\end{figure}

The simultaneous detection of two colors in single shots allows to observe the correlation between the pulse energies. An example of such a measurement for wavelengths 15 nm and 17 nm is presented in Fig.~\ref{spectr-corr}. One can observe, in fact, an anti-correlation which is a feature of the nonlinear regime of operation of the two-color scheme as explained above.

\begin{figure}[tb]
\includegraphics[width=.6\textwidth]{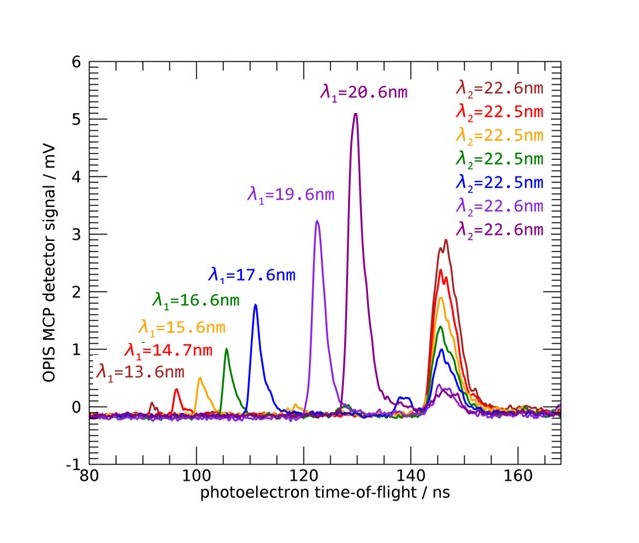}
\caption{2 color OPIS measurement while scanning color 1 from $\lambda_1$= 13.6 nm to 20.6 nm.  Color 2 was kept at $\lambda_2$= 22.6 nm. The time of flight directly translates into photon beam wavelength. }
\label{opis_scan}
\end{figure}

Continuous monitoring of spectral features is possible using the OPIS spectrometer.  Fig. \ref{opis_scan} shows a spectral measurement with the FLASH2 OPIS, where one of the colors was kept constant while the other one was scanned.
The measurement varied the first color from 13.6 nm to 20.6 nm in seven steps while the second wavelength was kept at 22.6 nm.  
We clearly observed the  increase in resulting photon pulse energy with longer first wavelengths $\lambda_1$.  This online monitoring is needed as influence of changing one wavelength to the other was observed.

\subsection{Measurements of temporal properties}

In order to gain information about the pulse duration of the two XUV pulses, we used the THz streaking technique \cite{Fruehling09, Grguras12}. THz streaking uses a noble gas target that is photoionized by the FEL pulses. The generated photoelectrons propagate in the time-varying electric field of the co-propagating THz radiation. After interaction with the terahertz field, the photoelectrons have changed their momentum component in the direction of the field. If the electron wave packet is short compared to the period length of the THz field, the temporal structure of the wave packet will be mapped onto the kinetic energy distribution of the emitted electrons and can be used to determine the pulse duration.

\begin{figure}[tb]

\includegraphics[width=.7\textwidth]{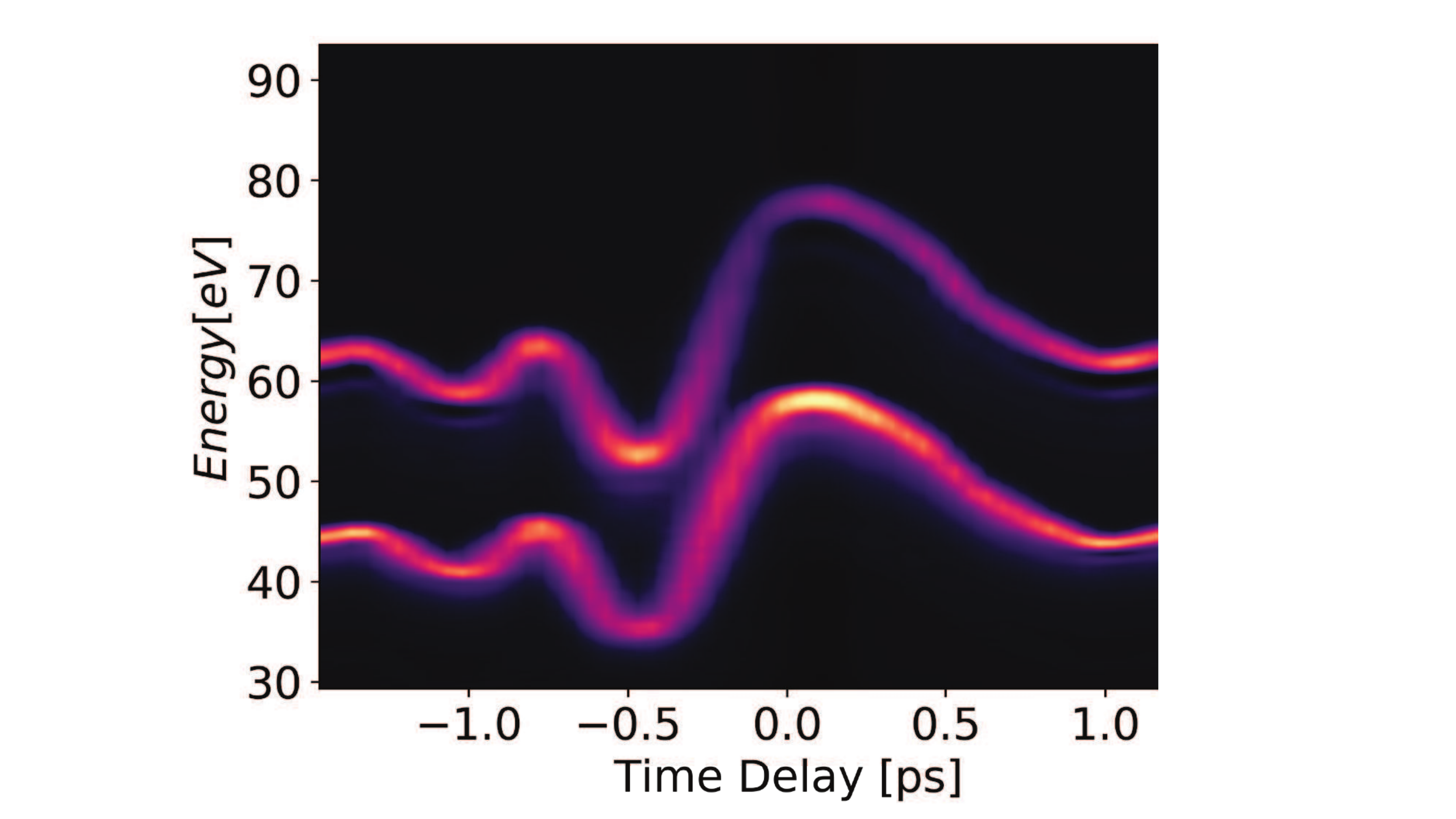}

\caption{The plot shows the vector potential of the THz streaking field recorded in a delay scan. The THz pulses were temporally scanned in respect to the XUV pulses. For each delay 30 photoelectron spectra were recorded. For the pulse duration analysis, the spectra in the middle of the linear slope (delay -0.2ps) were used. One can also see the broadening of the spectral lines in respect to the larger delays (with very low THz field), containing the information about the pulse duration.}

\label{streaking-scan}
\end{figure}

\begin{figure}[tb]

\includegraphics[width=.7\textwidth]{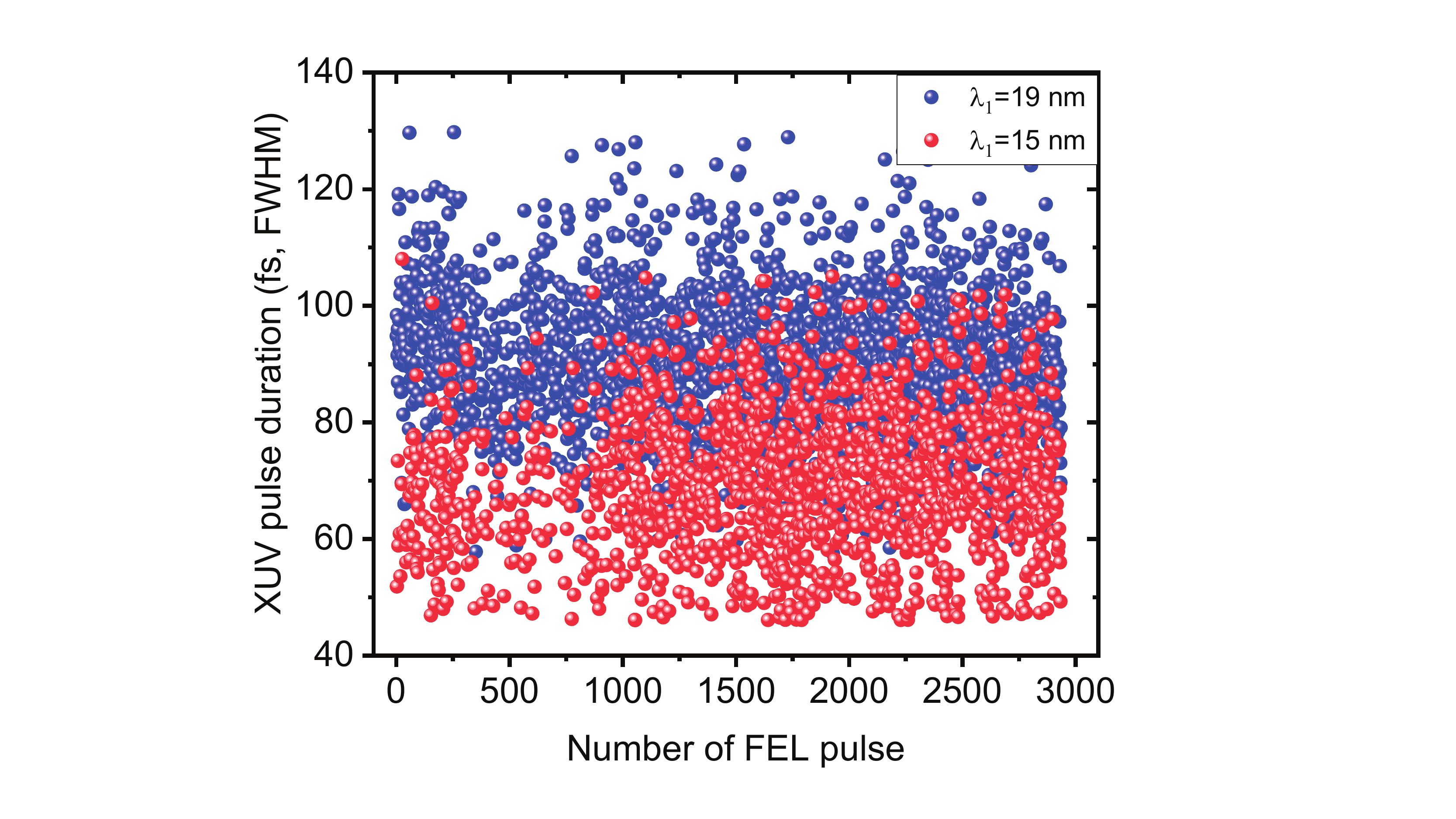}

\caption{Measured pulse duration for 3000 FEL pulses. The shorter wavelength $\lambda_1= 15$nm (red) shows a shorter pulse duration on average of 70 fs FWHM, while the pulse duration of $\lambda_2= 19$nm (blue) is 90 fs FWHM.}

\label{streaking-pulseduration}
\end{figure}

The measurements were performed at the dedicated photon diagnostic beamline FL21, which is equipped with a permanently installed THz streaking setup (for details see e.g. \cite{Ivanov18,Ivanov20}).
For the shown data, the FEL was operated at wavelengths of $\lambda_1$ = 15 nm (83 eV) and $\lambda_2$ = 19 nm (66 eV). Using Ne as target gas leads to kinetic energies of 61 and 45 eV for the 2p photoelectron lines, resulting from the two different wavelengths. Fig. \ref{streaking-scan} shows the kinetic energy of the two photoelectron lines as function of the delay between the THz field and the XUV pulse. This delay scan depicts the vector potential of the THz field and shows the linear slope which is used to map the temporal information into kinetic energy.
Acquiring data at a fixed delay in the middle of the linear slope we can determine the pulse-to-pulse variation of the pulse duration by monitoring the broadening of the streaked photoelectron line in respect to an unstreaked reference (THz off). The result for the two wavelength is shown in Fig. \ref{streaking-pulseduration}. The averaged pulse duration of $\lambda_1$ is 70$\pm$12 fs and 90$\pm$12 fs for the second wavelength, respectively. The given error denotes the RMS width of the distribution of the pulse durations.  Analyzing the shot-to-shot fluctuations it turns out that the pulse duration as well as the pulse energy (determined as integral over the photoelectron line) fluctuate independently for the two colors. No correlation was found, this is the consequence of the fact that FEL operated in linear regime. 

\begin{figure}[tb]

\includegraphics[width=.7\textwidth]{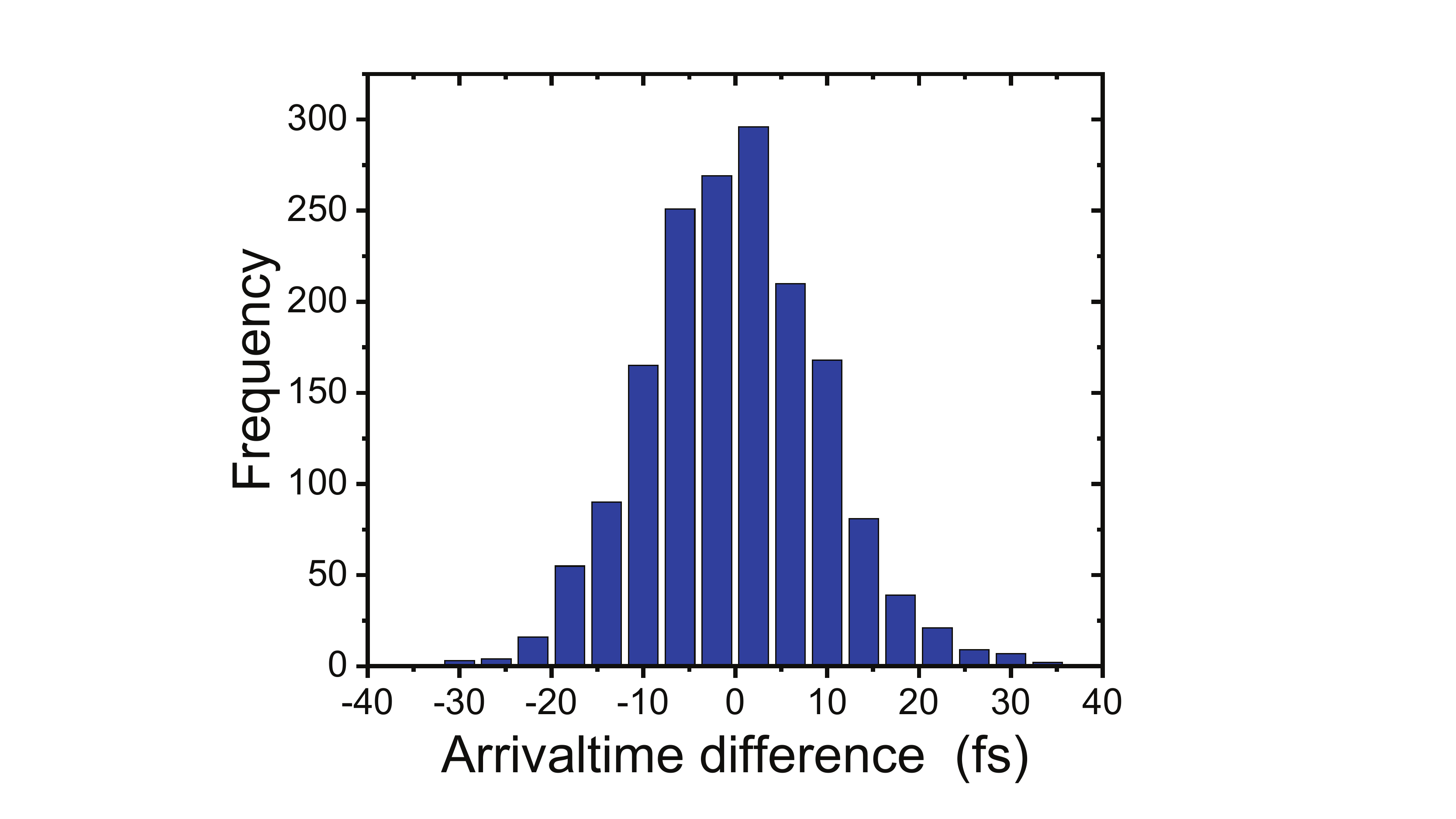}

\caption{Histogram of the relative arrival time fluctuations between the two pulses. For the used FEL parameters we found a $\sim$ 10 fs rms fluctuation of the relative arrival time. }

\label{streaking-arrivaltime}
\end{figure}

The method of THz streaking also allows to determine the arrival time of the XUV pulses in respect to the THz \cite{Ivanov18}. Here, the technique can be used to investigate the relative arrival times between the two-color pulses. First of all, a temporal shift between the two streaking curves in Fig. \ref{streaking-scan} indicates an average offset of the arrival time of the two pulses. For the shown experimental data we found a slight offset, indicating that the pulses with $\lambda_1$ arrive on average of $\sim$15 fs earlier as pulses with $\lambda_2$. Further measurements will be needed to investigate the dependence of the shift on FEL parameters.
Secondly, we can analyze the shot-to-shot arrival time fluctuations between the 2 pulses. As it was shown in Ref. \cite{Bermudez21} the SASE based fluctuations of the XUV pulse shapes lead to a variation of the mean arrival time (center of gravity), which is between 10 and 30\% of the pulse duration depending on the XUV pulse length and degree of saturation. For the measured settings we found that the relative temporal fluctuations between the two pulses follows a Gaussian distribution (see Fig. \ref{streaking-arrivaltime}) with an RMS width of 10 fs, in good agreement with the theoretical expectations of Ref. \cite{Bermudez21}.

Using the two color pulses for pump-probe experiments (e.g. using split and delay units) one has to be aware of the shot-to-shot fluctuation of the pulse duration as well as of a potential average offset. The effects are, however, significantly smaller as the pulse duration and will not lead to substantial loss in temporal resolution for experiments.

\section{Summary and outlook}

A method of two-color operation of an FEL with alternation of undulator tunes was successfully developed at the user facility FLASH. Parameters of the radiation (wavelength range and tunability, pulse energy, pulse duration, repetition rate) as well as developed photon diagnostics make such a source unique and suitable for different user experiments. In the future we plan to use a split-and-delay unit with filters so that pump-probe experiments with $\sim$ 10 fs resolution will become possible.

\section{Acknowledgement}
We would like to thank Wim Leemans and Edgar Weckert as well as scientific and technical staff of FLASH for support. We are grateful to Markus Ilchen for careful reading of the manuscript and useful comments. 
The work was carried out when Bart Faatz was still employed as a scientist at DESY.
This paper is dedicated to the memory of Andrey Sorokin, our colleague and friend.

\end{document}